# Back-action Evading Measurements of Nanomechanical Motion

J.B. Hertzberg[1,2], T. Rocheleau[1], T. Ndukum[1], M. Savva[1], A.A. Clerk[3], K.C. Schwab[4,**]

[1]*Laboratory for Atomic and Solid State Physics, Department of Physics, Cornell University, Ithaca, NY 14853 USA*

[2]*Department of Physics, University of Maryland, College Park, MD 20742 USA*

[3]*Department of Physics, McGill University, Montreal, H3A 2T8 CA*

[4]*Applied Physics, Caltech, Pasadena, CA 91125 USA*

**When performing continuous measurements of position with sensitivity approaching quantum mechanical limits, one must confront the fundamental effects of detector back-action. Back-action forces are responsible for the ultimate limit on continuous position detection, can also be harnessed to cool the observed structure [1,2,3,4], and are expected to generate quantum entanglement[5]. Back-action can also be evaded[6,7,11], allowing measurements with sensitivities that exceed the standard quantum limit, and potentially allowing for the generation of quantum squeezed states. We realize a device based on the parametric coupling between an ultra-low dissipation nanomechanical resonator ($Q \sim 10^6$) and a microwave resonator. [20] Here we demonstrate back-action evading (BAE) detection of a single quadrature of motion with sensitivity 4 times the quantum zero-point motion,**

$\Delta x_{ZP} = \sqrt{\dfrac{\hbar}{2m\omega_{NR}}}$ , **back-action cooling of the mechanical resonator to $\bar{n}_{NR} = 12$ quanta, and a parametric mechanical pre-amplification effect which is harnessed to achieve position resolution a factor 1.3 times $\Delta x_{ZP}$.**

When attempting to obtain complete knowledge of the dynamics of a simple harmonic oscillator, $x(t)$, back-action effects combined with the quantum zero-point motion of the oscillator limit the ultimate resolution to the standard quantum limit (SQL) **[8,9]**. The origin of this limit is the primitive fact that position and momentum are non-commuting observables, $[\hat{x}, \hat{p}] = i\hbar$, and are then linked through the equations of motion, $\dfrac{d\hat{x}}{dt} = -\dfrac{i}{\hbar}[\hat{x}, \hat{H}]$, where $\hat{H} = \dfrac{1}{2} m\omega_{NR}^2 \hat{x}^2 + \dfrac{1}{2m} \hat{p}^2$ is the system Hamiltonian of an oscillator with mass $m$ and natural frequency $\omega_{NR}$. However, during the theoretical investigations of the quantum limits of gravitational wave detectors over 30 years ago, it was realized that not all oscillator observables suffer from this fundamental limitation on measurement precision**[6, 7, 10 ]**. The two quadratures of motion $\hat{X}_1$, $\hat{X}_2$, where $\hat{x}(t) = \hat{X}_1 \cos(\omega_{NR} t) + \hat{X}_2 \sin(\omega_{NR} t)$, are non-commuting, $[\hat{X}_1, \hat{X}_2] = \dfrac{i\hbar}{2m\omega_{NR}}$, but are not linked dynamically and are constants of the motion: $\dfrac{d\hat{X}_i}{dt} = \dfrac{\partial \hat{X}_i}{\partial t} - \dfrac{i}{\hbar}[\hat{X}_i, \hat{H}] = 0$. Furthermore, it was realized how to couple $\hat{X}_1$ to a detector in a way which does not perturb the dynamics of $\hat{X}_1$: $\left[\hat{X}_1, H_{int}\right] = 0$, where $H_{int}$ is the interaction Hamiltonian. [12,13] Thus, while a measurement of $\hat{X}_1$ necessarily disturbs $\hat{X}_2$, this disturbance has no effect on the subsequent dynamics or measurements of X1. One can thus increase the coupling strength arbitrarily without fear of back-action, meaning that the sensitivity of such an ideal single quadrature probe is not fundamentally limited [11]. Such measurements are known as quantum non-demolition measurements (QND) [10]; the energy and phase of an oscillator form a similar pair of conjugate QND variables[6].

---

* Correspondence should be sent to schwab@caltech.edu

In the early 1980's it was realized that position detectors formed by the parametric coupling between electrical and mechanical oscillators can couple to a single mechanical quadrature of motion and can evade both quantum and excess classical sources of back-action [12,13]; excess classical back-action forces are created by noise injected into the electrical resonator by non-ideal following amplifiers. Furthermore, single-quadrature back-action evasion (BAE) techniques also are free from detector-induced mechanical damping [11,13].

These BAE techniques have been demonstrated on a number of gravitational wave detectors around the world[13], however they operate far from quantum mechanical limits. More recently, a BAE scheme utilizing the interference of two mechanical resonators in an optical cavity has been proposed[14] and been shown to partially evade classical back-action[15]. This scheme is however limited to a narrow, non-resonant frequency band, and does not allow squeezing or a true QND measurement. In this work, we demonstrate the manipulation of mechanical back-action effects of a stream of microwave photons. We realize a continuous, broad-band BAE scheme which allows resolution of a single quadrature near the zero-point motion, $\Delta x_{ZP}$, and offer a path to QND quadrature detection with sensitivity below $\Delta x_{ZP}$.

The approach we have taken utilizes a radio-frequency nanomechanical resonator (NR) which is coupled tightly to a microfabricated superconducting microwave resonator (SMR) [16], and stimulated with a stream of microwave photons, shown in Figure 1. Our nanomechanical resonator (NR) is formed from high-stress silicon nitride which shows very low dissipation rates[17] with $Q>10^6$ at a temperature of 15mK. Two devices were cooled in a dilution refrigerator and probed through carefully filtered, high-bandwidth cables. Using device 1 ($\omega_{NR} = 2\pi \times 5.57$ MHz, $\omega_{SMR} = 2\pi \times 5.01$ GHz), we studied backaction-evading measurement. Using device 2 ($\omega_{NR} = 2\pi \times 6.37$ MHz, $\omega_{SMR} = 2\pi \times 4.97$ GHz), we further investigated backaction cooling at high pump powers. Recently, a somewhat similar device demonstrated back-action cooling from N=700 to N=140[18,19] and continuous position detection a factor of $\Delta x = 30 \cdot \Delta x_{ZP}$ [20]; BAE techniques were not explored.

The expected quantum Hamiltonian of our parametrically coupled system is given by:

$$H = \hbar\left(\omega_{SMR} + g\hat{x} - \lambda\hat{x}^2\right)\left(b^+b + \frac{1}{2}\right) + \hbar\omega_{NR}\left(a^+a + \frac{1}{2}\right),$$ where $a$ ($a^+$) and $b$ ($b^+$) are the mechanical and electrical oscillator creation (annihilation) operators. The first term

shows the parametric coupling of the SMR's frequency to the mechanical motion: $\hat{x} = \Delta x_{ZP}(a^+ + a)$ and $g = \frac{\partial \omega_{SMR}}{\partial x} = \frac{\omega_{SMR}}{2C_T}\frac{\partial C_g}{\partial x}$ where $C_g(x)$ is the coupling capacitance and $C_T$ is the SMR's total effective capacitance. The term proportional to $\hat{x}^2$ results from the frequency pulling of the mechanical resonator by the SMR energy[21], where $\lambda = \frac{\omega_{SMR}}{2C}\frac{\partial^2 C_g}{\partial x^2}$. As we show below, this second-order term becomes important during BAE measurements.

When the SMR is pumped, harmonic motion of the NR modulates the SMR resonant frequency, causing pump photons to be both up-converted and down-converted by $\omega_{NR}$ [22, 23]. An up (down) converted photon is a result of the absorption (emission) of one mechanical quantum and leads to an increase (decrease) of NR damping and NR cooling (heating). The rate of up or down conversion is maximized when the final photon is degenerate with the cavity resonance: corresponding to a pump frequency of $\omega_{RED} = \omega_{SMR} - \omega_{NR}$ in the up conversion case, and $\omega_{BLUE} = \omega_{SMR} + \omega_{NR}$ in the down conversion case. When $\omega_{NR}/\kappa > 1$, the sideband resolved limit, this process can in principle cool the NR to the quantum ground state, $\bar{n}_{NR} = \langle a^+ a \rangle < 1$ [22,23], and is essential to form a BAE transducer[13]. Sideband resolved cooling of a parametrically coupled mechanical/microwave transducer was first demonstrated with a gravitational wave transducer [24], cooling from T=5K ($\bar{n} = 10^8$) to T=2mK ($\bar{n} = 10^5$), and has recently been explored electronically [18] and opto-mechanically [3,4].

Figure 2 shows the back-action effects on both the NR damping rate, $\Delta f_{NR}$, and thermal occupation factor, $\bar{n}_{NR}$, while driving the SMR with a single-tone pump, with $\omega = \omega_{RED}$ or $\omega = \omega_{BLUE}$. The sideband produced is a measure of both mechanical quadratures of the mechanical motion and subject to the usual SQL on position detection [13,25]. When applying the largest possible pump powers at $\omega = \omega_{RED}$, with $\bar{n}_{SMR} \sim 2 \cdot 10^8$, we find a position resolution of $\Delta x = 6.9 \cdot \Delta x_{ZP}$, by comparing the sideband amplitude to the noise floor. As the measurement power ($\bar{n}_{SMR}$) is increased the sensitivity approaches a limiting value of $\Delta x = 6.7 \cdot \Delta x_{ZP}$ due to the increase in back-action damping of the NR. This limiting resolution cannot be improved by improving NR properties or increased coupling to the SMR, and can only be improved with a superior microwave detector. The force sensitivity is limited by thermal motion of the NR to $1.7 \cdot 10^{-18} \, N/\sqrt{Hz}$ at the measurement temperature of 142 mK. The best achieved with this device was $8 \cdot 10^{-19} \, N/\sqrt{Hz}$ at 60 mK, equal to the highest force sensitivity achieved [1,26]. An essential ingredient to realize this position resolution is preparing highly monochromatic microwave photons. We use high-Q, cryogenic, tunable copper cavities to filter the phase noise of our microwave pumps, and achieve $\mathscr{L}$(+5.5MHz) = -195 dB$_c$/Hz, comparable to the best sources demonstrated [27].

Using device 2, we extend the backaction cooling to higher pump powers, achieving an occupation factor of $\bar{n}_{NR} = 12 \pm 4$, a factor of 2 below the lowest $\bar{n}_{NR}$ obtained using passive refrigeration, $\bar{n}_{NR} = 25$ [1]. Cooling in both devices is limited by excitation of the SMR and thermal heating of the NR at the highest pump powers, as well as a time varying, non-thermal dissipative force noise bath which appears to dominate over the thermal force noise at temperatures below ~60mK. Furthermore, subsequent measurements have shown that the NR experiences unexpected and as of yet, unexplained frequency drift and jitter comparable to $\Delta f_{NR}$ (~5-15 Hz) and variation of the damping rate on the history of the amplitude of mechanical motion and microwave drive.

To surpass the position sensitivity found when using a single cavity drive tone, we drive simultaneously with both red and blue pumps. Taking $\omega_{BLUE} - \omega_{RED} = 2\omega_{NR} + 600Hz$, we balance the rates of up and down conversion, producing no back-action damping while remaining sensitive to both quadratures of the mechanical motion. Figure 2 shows $\Delta f_{NR}$ and $\bar{n}_{NR}$ which are essentially independent of $\bar{n}_{SMR}$. The position resolution $\Delta x$ in this case continues to improve as $\bar{n}_{SMR}$ increases, limited in principle to $\Delta x_{ZP}$ by back-action force fluctuations. Here we achieve $\Delta x = 4.2 \cdot \Delta x_{ZP}$, limited by the power handling of the SMR and the noise floor of our microwave detection circuit. Given that $\bar{n}_{NR}$ does not systematically increase as the measurement strength is increased, we can place limits on the back-action force noise and find that $\sqrt{S_x \cdot S_F^{BA}} \leq 90 \cdot \hbar$.

To perform BAE, we need a measurement scheme sensitive to only a single mechanical quadrature. This is accomplished by pumping with equal intensity phase-coherent pumps at both $\omega_{RED}$ and $\omega_{BLUE}$, such that the up- and down-converted sidebands coherently interfere, $\omega_{BLUE} - \omega_{RED} = 2\omega_{NR}$. In this case, the measurement field is modulated at the NR resonant frequency: $E(t) = E_0 \cdot \cos(\omega_{SMR}t) \cdot \cos(\omega_{NR}t)$ **[13].** Figure 4 shows the phase sensitive nature of this detection scheme: by driving the mechanical resonator with a small resonant electrostatic force, we see that the detected signal at $\omega_{SMR}$ depends sinusoidally on the relative phase between the modulation of the cavity field, $\cos(\omega_{NR} t)$, and the resulting motion. In this way, the signals which are detected and amplified, do not capture all the information about the NR motion; one learns almost exclusively information regarding the $X_1$ quadrature, without observing $X_2$.

For pump strengths below our highest values, we observe that $\Delta f_{NR}$ and $\bar{n}_{NR}$ are essentially unchanged due to the balance between up and down conversion (Figure 2.) The highest sensitivity to one quadrature is: $\Delta X_1 = 4.1 \cdot \Delta x_{ZP}$, which is a limited by the additive noise of our microwave amplifier. While this sensitivity is comparable to that achieved with the previous two measurement schemes, the BAE scheme differs in being subject to no fundamental quantum limit on position resolution. An ideal amplifier would enable our device to achieve a position resolution below the SQL, $\Delta X_1 = 0.7 \cdot \Delta x_{ZP}$, by using the BAE scheme. We believe this is the first time that quadrature measurements of sufficiently high resolution have been realized which should in principle produce conditionally squeezed states**[11].**

To probe the BAE nature of this scheme we inject microwave frequency noise (generated by a chain of noisy amplifiers) into the SMR. The resulting cavity fluctuations will act as a classical source of back-action, and will heat the mechanics. As show in Fig. 4, when we pump the cavity with a single tone at $\omega_{RED}$, the back-action driven mechanical motion results in a signal which is out of phase with the fluctuating cavity voltage, squashing the output noise. However, when we pump with two tones in BAE configuration, no mechanical signature is seen in the output noise. This is a direct consequence of the BAE: the measured $X_1$ quadrature is unaffected by the cavity fluctuations. The expected effectiveness of the BAE technique is given by the ratio of backaction fluctuations in $X_2$ to those appearing in $X_1$, $\Delta X_2^2 / \Delta X_1^2 = 32(\omega_{NR}/\kappa)^2 \approx 4100$ **[11,13].** Averaging time and slow drifts of the NR frequency limit the resolution of $X_1$ flucutations. Estimating the maximum resolvable signal in the noise spectrum, we find a BAE effectiveness $\Delta X_2^2 / \Delta X_1^2$ of at least 82.

When the largest pump powers are applied in the BAE configuration, we find both linewidth narrowing and a dramatic increase in NR noise temperature. This is a result of a parametric amplification effect which originates from electrostatic frequency pulling of the NR proportional to the square of the SMR charge**[21]**. In the BAE configuration, the NR spring constant is modulated at $2\omega_{NR}$, which results in degenerate parametric amplification of the NR. When the size of the periodic NR frequency shift becomes comparable to $\Delta f_{NR}$, significant mechanical parametric amplification is observed. At our highest BAE drives, we have observed $\Delta f_{NR}$ narrowed to 2.1 Hz, and amplification factors of 11.6dB, which yields a measurement imprecision corresponding to $\Delta x = 1.3 \cdot \Delta x_{ZP}$. **[28]**.

This parametric effect and eventual instability limits the pump power and sensitivity of this BAE scheme. Analysis shows that the parametric amplification and de-amplification is in a basis which is rotated by π/4 from the measured quadrature $X_1$. Figure 3(b) shows this parametric gain at three different pump powers, observed by homodyne detection of the sideband of a weak red-detuned microwave probe signal, while driving the mechanics at $\omega_{NR}$. Although this effect is expected to destroy the BAE nature of the two tone detection, it offers other beneficial effects as it modifies the NR dissipation rate without associated noise. The increase in $\bar{n}_{NR}$ is a result of mechanical pre-amplification of the amplified quadrature. We have observed thermal noise squeezing[29] of the de-amplified quadrature which will be described in a future publication.

The techniques demonstrated here show the power of back-action engineering when performing strong measurement, and are expected to enable a number of significant advancements in the area of quantum state preparation and measurement of a mechanical device. The quantum ground state appears within reach with a combination of increasing SMR frequency, impedance, coupling $\partial C_g / \partial x$ and critical current of the SMR (e.g. using Nb-based SMR.) Furthermore, by increasing the SMR fundamental frequency to 10 GHz, while keeping all other device parameters the same, we expect to be able to realize sensitivity to one quadrature of $\Delta X_1 = 0.5 \cdot \Delta x_{ZP}$. This would allow the realization and study of squeezed mechanical states [11] which can be useful for ultra-sensitive detection and also provide a quantitative measurement of decoherence[30].

**Acknowledgement**: We would like to acknowledge helpful conversations with Gerard Milburn, Markus Aspelmeyer, Britton Plourde, Miles Blencowe and Roberto Onofrio, and the assistance of Peter Hauck, Madeleine Corbert, Sara Rothenthal and Chris Macklin. The work has been supported by Cornell University and grants from FQXi and the National Science Foundation. A.A.C. wishes to thank the Canadian Institute for Advanced Research. Device fabrication was performed at the NSF-sponsored Cornell Nanoscale Facility.

**Author contribution**: Authors J.B.H., T.R. and T.N. contributed equally to this research.

FIGURE 1

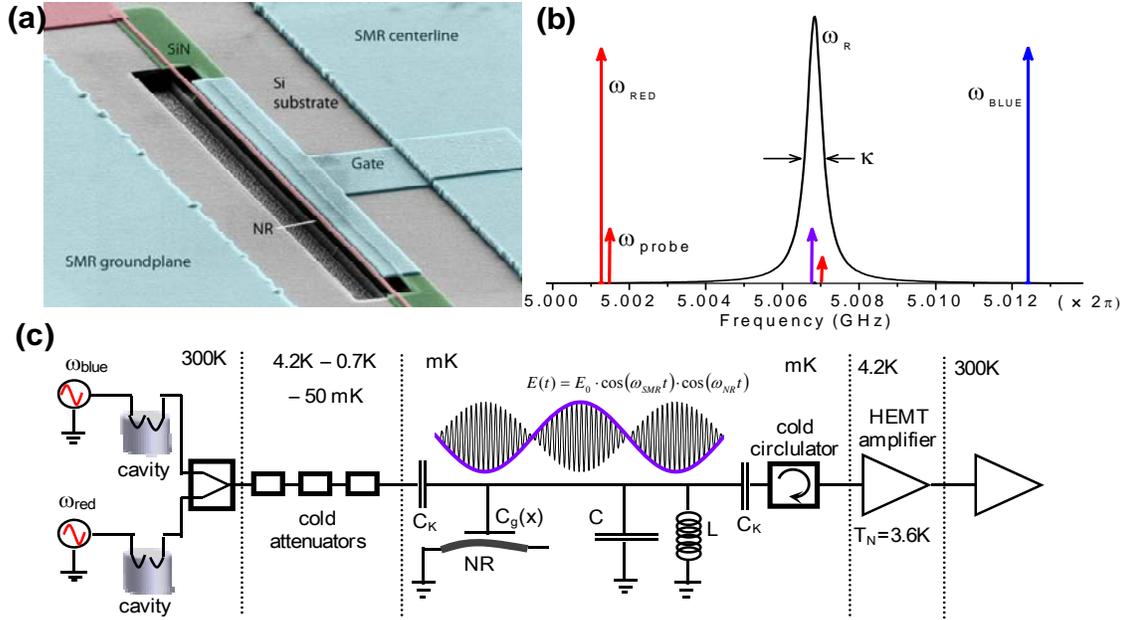

**FIGURE 1 | Device and measurement scheme.** (a) shows a false-colored SEM micrograph of device 1: a 260nm thick Al SMR, capacitively coupled ($C_g$, 85 nm gap) to an Al coated (105nm) high stress SiN (60nm thick) nanomechanical resonator (NR): 30µm x 170nm x 165nm. The SMR is formed by a 11.8 mm long, 50Ω co-planar wave guide (CPW) with a 16 µm wide center line, with matched ~4.5 fF coupling capacitors, on a high resistivity (>10kΩ) <100> Si wafer, with damping rate $\kappa = 2\pi \cdot 500$ kHz. Device 2 incorporates a 115Ω niobium CPW with $\kappa = 2\pi \cdot 272$ kHz but is otherwise similar to device 1. (b) shows a measurement of the device 1 SMR transmission and the spectral location of the microwave pumps and up- and down-converted photons. (c) shows a schematic of our cryogenic microwave measurement circuit. Total capacitance $C_T = C + C_g + 2C_K$ .

FIGURE 2

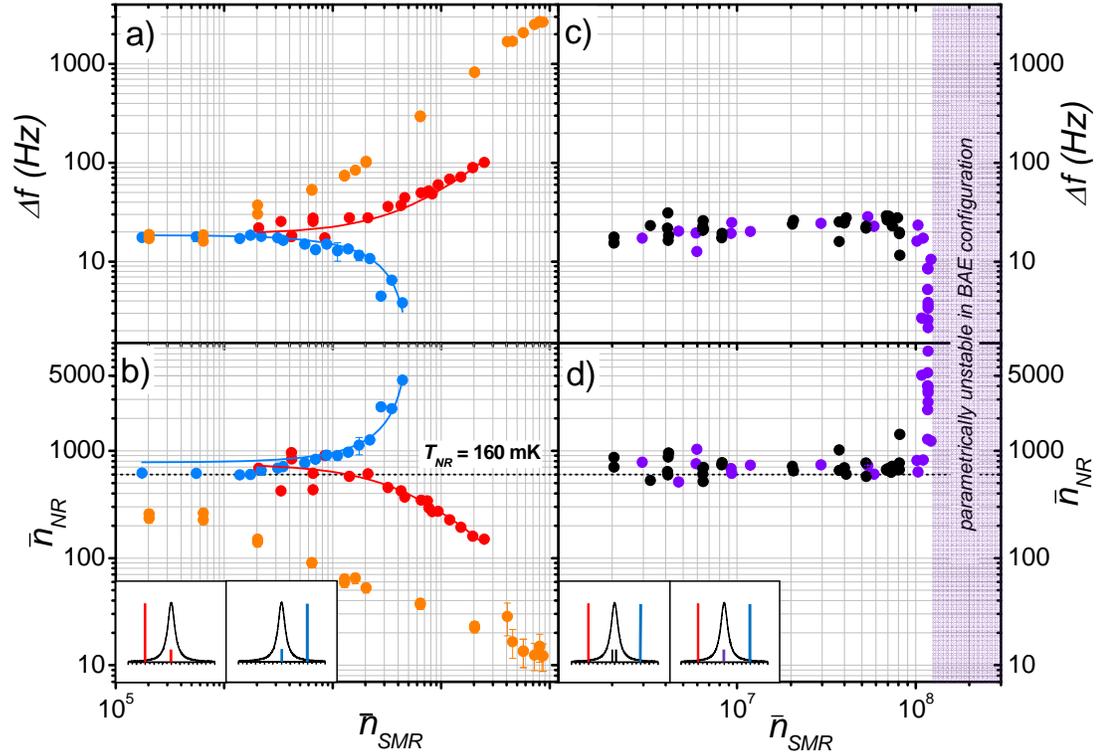

**FIGURE 2 | Linewidth, $\Delta f_{NR}$, and occupation factor, $\bar{n}_{NR}$, vs. SMR occupation. (a)** and **(b)** show NR behavior for a single pump tone: $\omega_{red}$ (●,●), or $\omega_{blue}$ (●). Solid lines show fits of the data to expressions in [11]. (See Supplementary Online Materials for details.) **(c)** and **(d)** show behavior for two pump tones: $\omega_{blue} - \omega_{red} = 2\omega_{NR} + 600 Hz$ (●), and $\omega_{blue} - \omega_{red} = 2\omega_{NR}$ (●). Only in the BAE configuration (●), at high pump power we observe $\Delta f_{NR}$ narrowing and mechanical amplification due to the parametric amplification. Drifts in the NR damping rate and frequency result in significant scatter in the (●) points undergoing parametric amplification. Shaded region is inaccessible to BAE

due to parametric instability. Insets show spectrally the arrangement of pump and sideband tones relative to the microwave resonance. (●,●,●,●) taken with device 1 at a fridge temperature of 142mK, (●) taken with device 2 at a fridge temperature of 100mK.

FIGURE 3

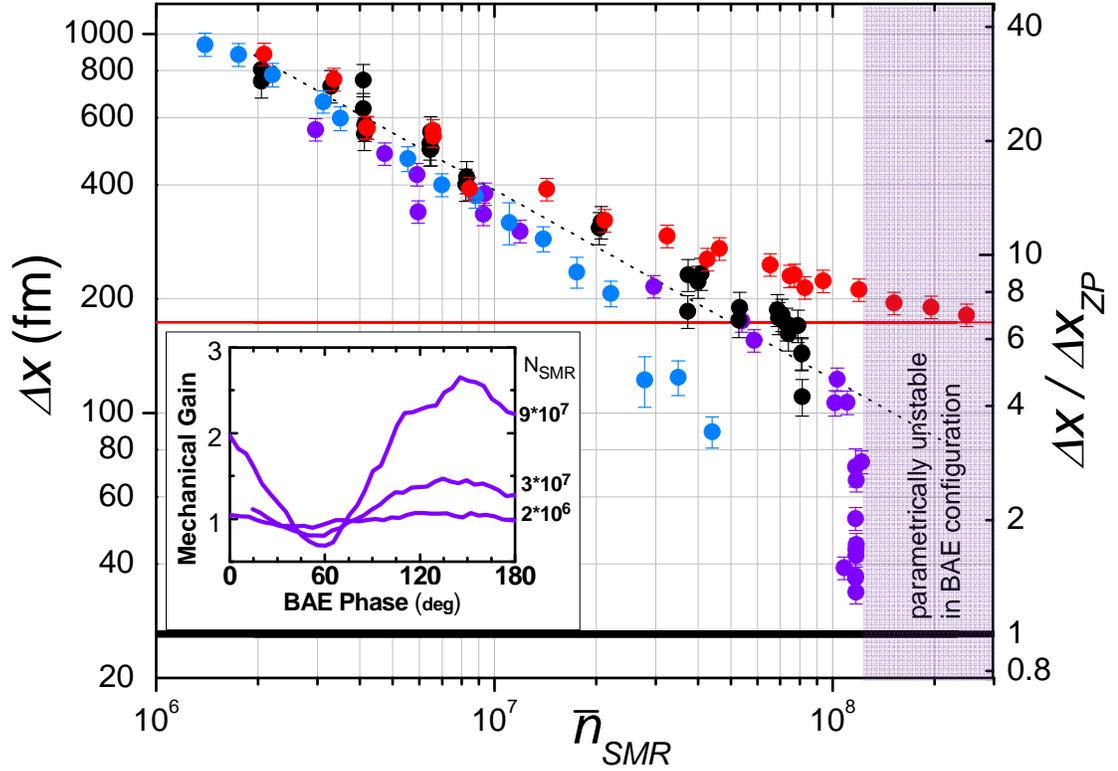

**FIGURE 3 | Measured position sensitivity vs. SMR occupation.** Device 1. Same pump configurations and symbols as used in Fig. 2. The horizontal red line shows the limiting sensitivity for a single pump tone. The slanted black line shows the expected sensitivity with no backaction damping. Shaded region is inaccessible to BAE due to parametric instability. The inset shows the parametric mechanical gain for three BAE pump levels, verses the SMR phase (referenced to the beating of the BAE pump.)

FIGURE 4

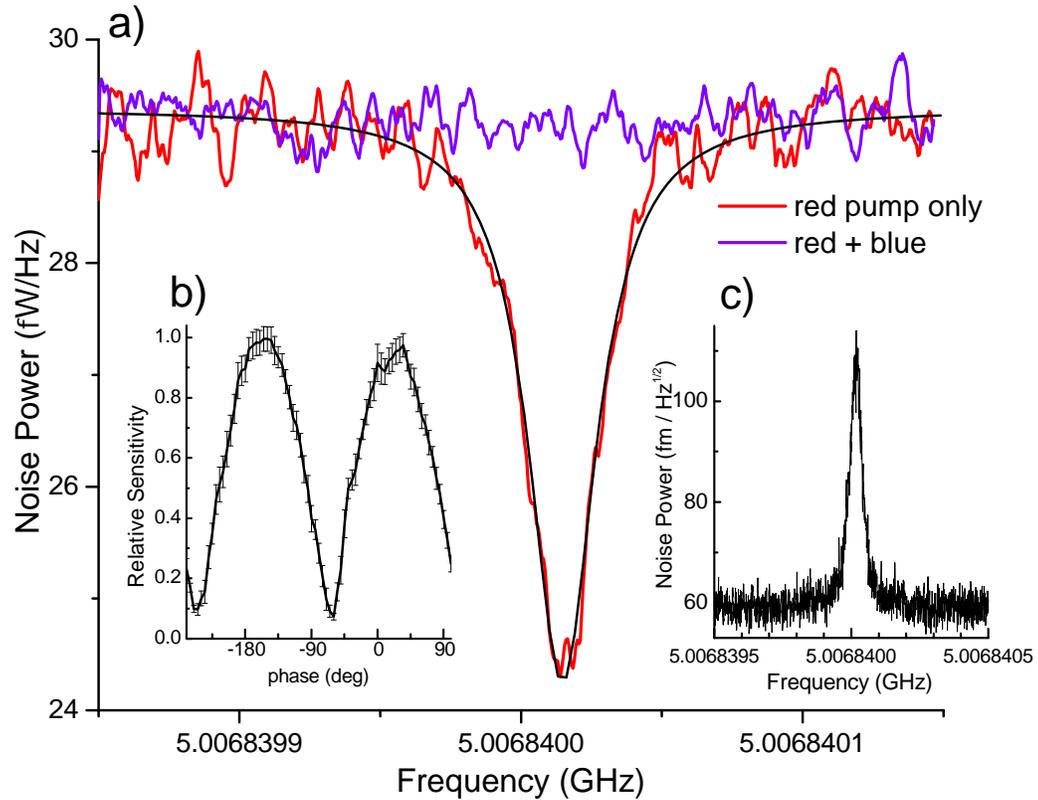

**FIGURE 4 | Single-quadrature detection and BAE.** Device 1. **(a)** shows the power spectrum at the SMR resonant frequency with 36 dB of excess noise injected to demonstrate the BAE scheme. Response to a single pump tone applied at $\omega_{RED}$ (red line data, black line lorentzian fit) shows a "hole" in the noise due to backaction-driven motion of the NR correlated with the SMR fluctuating voltage. Response to two pumps $\omega_{BLUE} - \omega_{RED} = 2\omega_{NR}$ (purple line) shows no backaction-driven motion in the measured quadrature. **(b)** shows the phase-sensitive nature of the scheme, coupling to only one mechanical quadrature. **(c)** shows the thermal motion of $X_1$ measured at 142mK demonstrating ultra-sensitive detection of the single quadrature.